\documentstyle[hip-artc]{article}
\volnumber{9}  \edyear{1999}  \frompage{000} \topage{000}                %|
\recrevdate{1 January 1999}                                              %|
\def\rightmark{ Pion entropy  and phase-space  densities...}
\def\leftmark{Yu.M. Sinyukov et al.}

\title{Pion entropy  and phase-space  densities  in  A+A  collisions}
\authors{
{\twerm Yu.M. Sinyukov$^1$ and S.V. Akkelin$^{1}$ %
}\\[2.812mm] {\normalsize \hspace*{-8pt}$^1$ Bogolyubov Institute
for Theoretical Physics, \\  03143 Kiev, Metrologichna 14b,
Ukraine\\ [0.2ex] }}

\abstract{We propose a method to estimate  the entropy of thermal
pions in A+A collisions  irrespective of unknown form of
freeze-out (isothermal) hypersurface and transverse  flows
developed. We analyse the average phase-space  densities and
entropies of the thermal pions vs their  multiplicities  and
collision energies.  The behaviour of these  values apparently
indicates the deconfinement and chiral  phase transitions in
relativistic nuclear-nucleus collisions.} 

\begin{document}

\maketitle

\section{Introduction}

The actual numbers of partonic degrees of freedom released in
ultra-relativistic nuclear-nucleus collisions are fairly big, up
to tens of thousands. As it is commonly supposed such a
quasi-macroscopic system becomes thermal at the early stage of
collision process: current phenomenological and pQCD estimates
give the times of thermalization between $0.6$ and 3 fm/c
\cite{time1}. Then the system expands nearly isoentropically -
this standard point in the hydrodynamic approach to A+A collisions
is advocated theoretically in the Refs. \cite{Shuryak}. Thereby
the entropy produced in A+A collisions carries out the information
about very initial thermal stage of these processes. It is found
recently \cite {AkkSin} that the pion phase-space density averaged
over momentum (totally or at fixed rapidity) and space is also
about the constant during the stage of the chemically frozen
expansion. So, measurements of the entropy and average phase-space
density (APSD) in thermal hadronic systems at the final
(freeze-out) stage of A+A collisions gives the possibility to look
inside the previous stages such as the partonic thermalization and
the hadronization (or chemical freeze-out). The aim of this work
is to estimate the entropy and the APSD of thermal pions in heavy
ion collisions at the different energies of  CERN SPS and BNL RHIC
and to analyse  the physical meaning of the results.

\section{The extracting entropy of thermal pions}

As well known the entropy of a gas of bosons/fermions  has the following
covariant form
\begin{equation}
\label{ent-def}S=(2J+1)\int \frac{d\sigma ^\mu p_\mu d^3{p}}{(2\pi )^3p^0}%
\,[-(2\pi )^3f\ln ((2\pi )^3f)\pm (1\pm (2\pi )^3f)\ln (1\pm (2\pi )^3f)]
\end{equation}
where the $\pm $ sign corresponds to bosons/fermions with total
spin $J$ and $\sigma $ is some hypersurface in Minkowski space.
The value depends on the distribution function, or phase-space
density $f(x,p)$ that should be known to make the correspondent
estimates. It is easy to show, however, that the phase-space
density, e.g. of $\pi ^{-}$, cannot be extracted from  two- and
many- particle spectra even if the system at the kinetic
freeze-out is characterized by the locally equilibrium
distribution function. To study the problem let us write the
Wigner function (the analogy of the phase-space density for
quantum systems) for weekly interacting particles in the
mass-shell approximation \cite{Groot}:
\begin{equation}
\label{Wigner1}f(x,p)=(2\pi )^{-3}\int d^4q\delta (q\cdot
p)e^{-iqx}\left\langle a^{+}(p-(1/2)q)a(p+(1/2)q)\right\rangle .
\end{equation}
Here the brackets $\left\langle ...\right\rangle $ mean the averaging of the
product of creation and annihilation operators with density matrix refereed
to a space-like hypersurfaces where particles become or are already nearly
free. If the freeze-out is sudden, one uses usually a thermal density matrix
at the freeze-out hypersurface. The invariant single- and double- particle
spectra have the form:
\begin{equation}
\label{spectra1}n(p)=\left\langle a^{+}(p)a(p)\right\rangle
,n(p_1,p_2)=n(p_1)n(p_2)+\left| \left\langle a^{+}(p_1)a(p_2)\right\rangle
\right| ^2,
\end{equation}
and the correlation function is defined as the following
\begin{equation}
\label{corr-definition}C(p,q)=n(p_1,p_2)/n(p_1)n(p_2).
\end{equation}
It is easy to see now that the phase of the two- operator average $%
\left\langle a^{+}(p_1)a(p_2)\right\rangle $ cannot be extracted
from the  single- and double- particle spectra. It is possible to
show that the same takes place even when many particle spectra are
included into an analysis. Therefore one cannot reconstruct the
distribution function $f(x,p)$ in a model independent way.
However, since the $f(x,p)$ is real (note, that for the (locally)
equilibrited quantum systems the Wigner function is positive), one
can use Eq. (\ref{Wigner1}) to express the $f^2(x,p)$ integrated
over space coordinates just through the squared absolute value of
the two- operator average, $\left| \left\langle
a^{+}(p-(1/2)q)a(p+(1/2)q)\right\rangle \right| ^2$,  integrated
over $q$. Then, accounting for the direct links (\ref{spectra1}),
(\ref {corr-definition}) of the later value with the single
particle spectrum (that is just the $f(x,p)$ integrated over space
coordinates) and correlation function one can extract the
phase-space density averaged over some hypersurface $\sigma $ at
or ''after'' freeze-out
\begin{equation}
\label{covariant}\left\langle f(\sigma ,p)\right\rangle =\frac{\int \left(
f(x,p)\right) ^2p^\mu d\sigma _\mu }{\int f(x,p)p^\mu d\sigma _\mu }=\frac
1{(2\pi )^3}\int \frac{n(p)}{p^0}(C(p,q)-1)d^3q
\end{equation}
directly from the experimental data in full accordance with the pioneer
Bertsch idea \cite{Bertsch}.

At the first sight the extracted value of the APSD does not help
to calculate the entropy (\ref{ent-def}), and so, some
phenomenological functions that reproduce the approximate Gaussian
behaviour of the correlation function are utilized usually
\cite{Pratt}. One should understand, however, that, since we
cannot extract the phase of the two- operator average, there are
an infinite set of distribution functions compatible with the
observables and, therefore, the entropy calculated will depend on
the class of the functions we choose which. Here we propose the
method how to estimate the entropy using just the APSD, do not
supposing any concrete expression for the phase-space density.

The method is similar to what was proposed in Ref. \cite{AkkSin}
to estimate the overpopulation of the phase-space. The idea is
based on the standard approach for spectra formation \cite{Landau}
that supposes the momentum hadronic spectrum in expanding locally
equilibrated system becomes frozen at some space-time hypersurface
with uniform temperature and particle number density. Then, within
this approximation which is probably appropriate in some
''boost-invariant'' midrapidity interval, the phase-space density
totally averaged over freeze-out hypersurface and momentum
excepting for the longitudinal one (rapidity is fixed, e.g.,
$y=0$) will be the same as the totally averaged phase-space
density in the static homogeneous Bose gas:
\begin{equation}
\label{result1}(2\pi )^3\left\langle f(\sigma ,y)\right\rangle _{y=0}=\frac{%
\int d^3p\overline{f}_{eq}^2}{\int d^3p\overline{f}_{eq}}
\end{equation}
where
\begin{equation}
\label{def1}\overline{f}_{eq}\equiv (\exp (\beta (p_0-\mu )-1)^{-1}
\end{equation}
and $\beta $ and $\mu $ coincide  with  the inverse of the
temperature and chemical potential at the freeze-out hypersurface
\cite{AkkSin}. The essence of the result is that under above
conditions any function of the locally-equilibrium distribution,
$F(f_{l.eq.})$, integrated with the measure $p^\mu d\sigma _\mu
d^3p/p^0$ contains the common factor, ''effective volume'' $\int
d\sigma _\mu u^\mu $,
that completely absorbs the flows $u^\mu (x)$ and form of hypersurface $%
\sigma (x)$. This factorization property has been found first in
Ref. \cite{Nukleonika}, multiple used for an analysis of particle
number ratios (see, e.g., Ref. \cite{A-B-S}) and recently of the
APSD in Ref. \cite{AkkSin}.

In this work we apply the approach to an analysis of the thermal pion
entropy per particle, or specific entropy of thermal pions. Using the same
approximation of the uniform freeze-out temperature and density and Eq. (\ref
{ent-def}) with local equilibrium distribution we get the following
expression for specific entropy in midrapidity:
\begin{equation}
\label{en-to-n}\frac{dS/dy}{dN/dy}= \frac{\int d^3{p}\,[-\overline{f}%
_{eq}\ln \overline{f}_{eq} + (1+ \overline{f}_{eq})\ln (1+
\overline{f}_{eq})] }{\int d^{3}p \overline{f}_{eq}}
\end{equation}
In the above ratio due to the factorization property the effective
volume is canceled and the final expression depends only on the
two parameters: temperature and chemical potential at the
freeze-out. The temperature can be obtained from the fit of the
transverse spectra for {\it different} particle species and we
will use the value $T=120$  MeV as the typical ''average'' value
for the SPS and RHIC experiments. The another parameter, chemical
potential, cannot be extracted from the spectra, its value could
be fairly high even for the thermal pions because of the chemical
freeze-out and this parameter is crucial for the entropy
estimates. We will extract it from the APSD analysis following to
 Ref. \cite{AkkSin}. First note that similar to Eq.
(\ref{covariant}) for $\left\langle f(\sigma ,p)\right\rangle $
one can express the totally averaged in midrapidity APSD
$\left\langle f(\sigma ,y)\right\rangle _{y=0}$ by means of the
transverse spectrum and correlation function in standard
Bertsch-Pratt parameterization \cite{Bertsch,AkkSin}:
\begin{equation}
\label{PSDA_exp}(2\pi )^3\left\langle f(\sigma ,y)\right\rangle
_{y\simeq 0} \approx
\kappa \frac{2\pi ^{5/2}\int \left( \frac 1{R_OR_SR_L}\left( \frac{d^2N}{%
2\pi m_Tdm_Tdy}\right) ^2\right) dm_T}{dN/dy}.
\end{equation}
The factor $\kappa $ is introduced to eliminate contribution of
short-lived resonances to the spectra and interferometry radii \cite{AkkSin}%
. It absorbs also the effect of suppression of the correlation
function due to  the long-lived resonances. Because of the
chemical freeze-out a big part of pions, about a half, are
produced by the  short-lived resonances after thermal freeze-out,
that leads to an additional non-equilibrium production of the
entropy. To eliminate these non-thermal contributions to the pion
spectra and correlation functions, we utilize  the detailed study
of the correspondent contributions presented in  Ref.
\cite{AkkSin} where within hydrodynamic approach it  was found
that $\kappa =0.65$ for SPS and $\kappa =0.7$ for RHIC if half of
pions is produced by resonances at the post freeze-out stage.
Then, by means of Eqs. (\ref{result1}), (\ref{PSDA_exp}) one can
extract the pion chemical potential. The latter value gives
possibility to estimate the phase-space overpopulation and the
specific entropy of thermal pions as it was explained above.

\section{Specific entropy and APSD at SPS and RHIC.}

To evaluate the APSD  of direct pions (\ref{PSDA_exp}) we utilize
the following parametrization of the $\pi ^{-}$ transverse spectra
and interferometry radii.

For SPS Pb+Pb(Au) 40,  80, and 158 AGeV ($\sqrt{s_{NN}}=$8.8, 12.3
and 17.3 GeV):

The transverse spectra are $\frac{d^{2}N}{2\pi
m_{T}dm_{T}dy}=A\exp (-m_{T}/T_{eff})$,
$T_{eff}\approx$ 0.169, 0.179, 0.180 GeV, the midrapidity densities are $\frac{dN_{\pi^{-}}%
}{dy}\approx$ 106.1, 140.4, 175.4 (NA49 Collaboration,
\cite{NA49}).
 The interferometry radii are  $R_{L}=C_{L}/\sqrt{%
m_{T}}$,   $R_{S} = $ \newline $ C_{S,1}/\sqrt{1+C_{S,2}m_{T}}$
and correspondent numerical parameters are taken from Ref.
\cite{CERES} (CERES Collaboration). We use approximation
$R_{O}=R_{S}$  until minimal measured  $p_{T}$ momentum,
$p_{T}=0.125$ GeV,   and our analytical approximation of the CERES
outward interferometry radii data for $p_{T}>0.125$ GeV.

For RHIC Au+Au $\sqrt{s_{NN}}=130$ GeV:

The transverse spectrum is $\frac{d^{2}N}{2\pi m_{T}dm_{T}dy}=A$
$(\exp (-m_{T}/T_{eff})-1)^{-1}$, $T_{eff}\approx 0.218$ GeV, and
$\frac{dN_{\pi^{-}} }{dy}\approx 249$ (STAR Collaboration,
\cite{STAR-N}). We use here the Bose-Einstein parameterization of
the transverse spectra from Ref. \cite{STAR-N} since the
integrated rapidity density of negative pions with this fitting
function is more close to presented recently by the PHENIX
Collaboration value $\frac{dN_{\pi^{-}} }{dy}\approx 270$  at
midrapidity \cite{phenix-new} than what one can get from the
exponential parameterization of the  transverse spectra,
$\frac{dN_{\pi^{-}} }{dy}\approx 229$ \cite{STAR-N}.
 The phenomenological parameterization of
interferometry radii, $R_{L}=C_{L}/ \sqrt{m_{T}}$,
$R_{S}=C_{S,1}/\sqrt{1+C_{S,2}+C_{S,3}m_{T}}$,  and the
correspondent numerical parameters are taken from Ref. \cite{STAR}
of the PHENIX Collaboration. As one can see from Fig. $3$ of Ref.
\cite{STAR}, there is some discrepancy  in  the data on $R_{S}$
radii between the STAR \cite{st-pion} and PHENIX Collaborations
\cite{STAR}. To optimize uncertainties in the forthcoming
estimates , we  utilize for the  parameter  $C_{S,1}$ the value
$8.75$ fm that is the average between "PHENIX data motivated"
value, $8.1$ fm, and "STAR data motivated" value, $9.4$ fm,
\cite{STAR}. We use here approximation $R_{O}=R_{S}$.

The results of our calculations of the APSD according to
(\ref{PSDA_exp}) are used then to extract the pion chemical
potentials and then the specific entropies basing on Eqs.
(\ref{result1}) and (\ref{en-to-n}) and supposing  the temperature
of thermal freeze-out $T_{th}=120$ MeV. The values founded are
collected in Table 1 and  in Figs.  1, 2,  3. We present in Table
$1$  also the freeze-out densities of thermal pions at different
SPS and RHIC energies.  The correspondence between points and
energies in  Figs.  1, 2, 3  one can  see  from Table 1.
\begin{center}
 TABLE 1

\bigskip

\begin{tabular}{|c|c|c|c|c|}
  \hline
$\sqrt{s_{NN}}$ & $8.8$ GeV & $12.3$ GeV & $17.3$ GeV & $130$ GeV
\\
  \hline
  $dN^{\pi^{-}}_{th}/dy$ & $106.1/2$  & $140.4/2$  & $175.4/2$  & $249/2$ \\
  \hline
$(2\pi )^{3}\left\langle f(\sigma ,y)\right\rangle^{\pi^{-}}_{th}$
&$0.125$ & $0.127$ & $0.137$& $0.168$ \\
  \hline
$\mu_{\pi}$ & $30.5$ MeV & $31.5$ MeV  & $37.7$ MeV  & $54$ MeV \\
\hline
 $dS^{\pi^{-}}_{th}/dy$ & $213.8$  & $282.2$   & $347.3$   & $471$  \\
 \hline
 $\frac{dS^{\pi^{-}}_{th}/dy}{dN^{\pi^{-}}_{th}/dy}$ & $4.03$  & $4.02$   & $3.96$   &
  $3.78$  \\
 \hline
  $n_{\pi^{-}}$ & $0.025$ fm$^{-3}$  & $0.025$ fm$^{-3}$   & $0.027$ fm$^{-3}$ &  $0.031$ fm$^{-3}$ \\
 \hline
\end{tabular}

\end{center}

\section{Discussion and conclusions}

Here we represent brief analysis and interpretation of the results
obtained. As was shown in Ref. \cite{AkkSin} the pion APSD
$\left\langle f(\sigma ,y)\right\rangle$ is approximately
preserved during isoentropic and {\it  chemically frozen}
expansion unlike fast decrease of the particle $n(x)$ and
phase-space $f(x,p)$ densities. Thereby the systems reach the
kinetic freeze-out at some typical particle densities that are
similar for different collision energies,  and, at the same time,
their  APSD increases with energy  since  it "memorizes" the
higher initial densities at higher energies. The rise with energy
of the pion APSD at AGS is seen from the behaviour of the
interferometry radii presented in Ref. \cite{CERES}: the
correspondent interferometry volumes $R_{O}R_{S}R_{L}$ do not
change or even  drop with energy while the pion rapidity densities
grow up. From Fig. 1 one can see  the increase of pion APSD also
from SPS to RHIC   energies. Only at low SPS energies
\newpage

\setlength{\textfloatsep}{10pt}

\begin{figure}[bhp]
\vspace*{15 mm}
\begin{minipage}[t]{5.5cm}
{\epsfysize=5.5 cm \epsfxsize=5.5cm \epsfbox{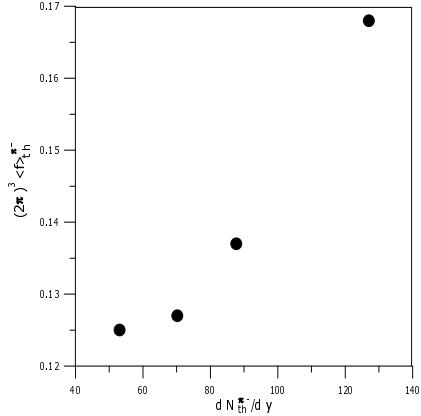}}
\end{minipage} \hfill
\vspace*{-40mm} \hspace*{6.5cm} \hfill
\begin{minipage}[t]{6.5cm}
\vspace*{2pt} \caption[]{\small  The average phase-space densities
of negative thermal pions $(2\pi )^{3}\left\langle f(\sigma
,y)\right\rangle^{\pi^{-}}_{th}$ as function of rapidity density
of negative thermal pions   $dN^{\pi^{-}}_{th}/dy$  at SPS and
RHIC energies.  Table 1 gives correspondence between  points  and
collision energies.} \label{fig1}
\end{minipage}
\end{figure}

\noindent the APSD values are approximately constant. The reason
for   the  latter could be the deconfinment happens in that
interval of energies. Then one should take into  account  a
relatively big gap at that energies between the temperatures of
the phase  transition (hadronization) and chemical freeze-out
\cite{PBM} when the hadronic system  evolves as {\it  chemically
equlibrated}. While the pion APSD might be fairly high at the
hadronization stage, it drops during the chemically equilibrated
stage of the evolution until the system reaches the chemical
freeze-out. At higher energies the temperature gap between
hadronization and chemical freeze-out becomes narrow \cite{PBM}
and so the pion APSD becomes  frozen and observed value tends to
what was at the hadronization stage, see points for the SPS
highest energy and for the RHIC one in Fig. 1.

The Fig. 2  demonstrates the linear increase of the entropy of
thermal negative pions with their rapidity density that can be
described approximately as $dS_{th}/dy=28+3.6*dN_{th}/dy$. Thereby
the specific entropy approaches the massless limit,
$S/N\rightarrow 3.6$ at $m/T\rightarrow 0$, when rapidity density
grows, see Fig. 3. The direct interpretation might be that the
chemically equilibrium pion number is frozen at about initial very
high temperature that increases with the collision energy. The
existence of the pions in the quark gluon plasma up to zero
binding temperatures $T\approx 1.6T_{c}$ is argued  recently
\cite{Shuryak}. However, the mass of pions in the QGP most
probably grows with temperature (see, e.g.,
\cite{Randrup,Shuryak}) that turns down the possibility of such an
explanation. Formally, the decrease of specific entropy of thermal
pions  is  conditioned by an increase  of the  pion chemical
potential at kinetic  freeze-out. The  latter  happens since  the
APSD  is conserved during the evolution  and it grows  with energy
at  the  chemical freeze-out.  Even if the thermodynamic
parameters at the chemical freeze-out do not change, the  APSD
grows if the intensity of flows increases, -  it is determined by
the correspondent reduction of the "interferometry volume"
\cite{AkkSin}. Since there is the strong correlation  between  the
initial densities in A+A collisions and intensity of the flow
developed, the conservation of the APSD results in an increasing
of observed of the pion chemical potential and an decrease of
specific entropy with collision energy if  the hadronization and
chemical freeze-out are nearly coincided.

\setlength{\textfloatsep}{10pt}

\begin{figure}[bhp]
\vspace*{10mm}
\begin{minipage}[t]{5.5cm}
{\epsfysize=5.5 cm \epsfxsize=5.5cm \epsfbox{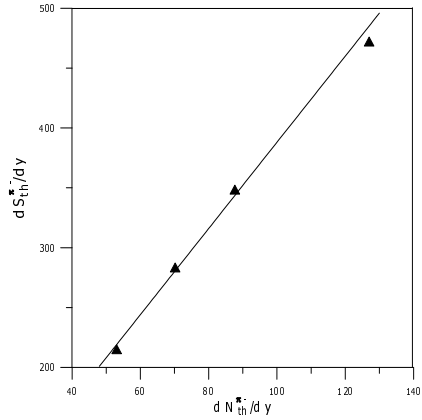}}
\end{minipage} \hfill
\vspace*{-40mm} \hspace*{6.5cm} \hfill
\begin{minipage}[t]{6.5cm}
\vspace*{2pt} \caption[]{\small  The entropy rapidity density of
negative thermal pions   $dS^{\pi^{-}}_{th}/dy$ as function of
rapidity density of negative thermal pions $dN^{\pi^{-}}_{th}/dy$
at SPS and RHIC energies. The  solid line corresponds to linear
approximation $28+3.6*dN^{\pi^{-}}_{th}/dy$.} \label{fig2}
\end{minipage}
\end{figure}

\setlength{\textfloatsep}{10pt}

\begin{figure}[bhp]
\vspace*{10mm}
\begin{minipage}[t]{5.5cm}
{\epsfysize=5.5 cm \epsfxsize=5.5cm \epsfbox{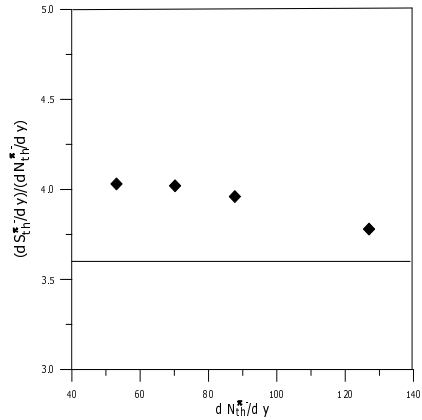}}
\end{minipage} \hfill
\vspace*{-40mm} \hspace*{6.5cm} \hfill
\begin{minipage}[t]{6.5cm}
\vspace*{2pt} \caption[]{\small The specific entropy of negative
thermal pions  $\frac{dS^{\pi^{-}}_{th}/dy}{dN^{\pi^{-}}_{th}/dy}$
as function of rapidity density of negative thermal pions
$dN^{\pi^{-}}_{th}/dy$  at SPS and RHIC energies. The  solid line
corresponds to  specific  entropy in massless  gas.} \label{fig3}
\end{minipage}
\end{figure}

\noindent

Some  peculiarities of the energy  dependence of specific entropy
could be  associated also with the deficit  of pions  at highest
SPS  and probably RHIC energies if one supposes pion chemical
equilibrium at chemical freeze-out \cite{A-B-S}. The necessary
additional contribution can be associated with decays of $\sigma$
mesons \cite{Koch}. Due to smaller mass of $\sigma$ at the phase
boundary as compare to its vacuum mass, the chiral phase
transition is accompanied by an appearance  and subsequent decay
at post hadronization stage  of a big number of these particles.
Therefore, besides the basic increase  of the pion APSD with
energy at the chemical freeze-out, the above mechanism can lead to
an increase of the APSD at subsequent evolution and, thus,  to
additional decrease of specific entropy of thermal pions.

\section*{Acknowledgments}

The work was supported by NATO Collaborative Linkage Grant No.
PST.CLG.980086.

\vfill\eject
\end{document}